# Measuring entanglement in material traces of ritualized interaction: Preferential attachment in a prehistoric petroglyph distribution

Tom Froese [1, *] and Emiliano Gallaga [2]

[1] Embodied Cognitive Science Unit, Okinawa Institute of Science and Technology Graduate University, Okinawa, Japan
[2] Universidad Autónoma de Chiapas, Mexico

* Corresponding author's e-mail: tom.froese@oist.jp

**Abstract**

Prehistoric rock art is often analyzed predominantly as the product of artists' intentions to create public representations of their perceptual experiences and mental imagery. However, this representation-centered approach tends to overlook the performative role of much material engagement. Many forms of rock art are better conceived of as traces from artists' repeated engagement with a surface, including with previous traces. For these artists, a potentially more relevant intention was ritualized interaction, such as communion and petition, which were realized as materially mediated transactions with the agencies that were believed to animate specific areas of the environment. If so, we can expect the motifs to be strongly clustered on ritually attractive areas, rather than to be evenly distributed on canvas-like surfaces that would maximize their visibility as public representations. Here we propose a novel way of testing the interaction-centered approach in terms of preferential attachment, which is a concept from network science that describe the well-known social phenomenon that popular agents tend to attract more followers. We applied this approach to a case study of an archaic site in Chihuahua, Mexico, and found that its petroglyph distribution has the form of a power law, which is consistent with preferential attachment. We conclude that this approach could be developed into a measure of the entanglement between ritual processes and products in prehistoric material engagement.

**Introduction**

Rock art distributions are often puzzling. Imagine you are part of an archaeological survey expedition to record petroglyphs in a large desert valley. You are on your way to a rock outcrop that once could have served as a highly visible landmark, and which you know from aerial photos was also in an area that served as a source of water. As you arrive with the morning sun rising into the sky behind you, you notice that a couple of its boulders are dotted with a few petroglyphs (**Figure 1B**). But as you walk around to the other side, you suddenly find yourself confronted with a large number of rock art panels, including a few with exceptionally dense concentrations of petroglyphs, to the point that they overlap and occlude each other (**Figure 1A**).

Why did their creators not take advantage of all the highly suitable surfaces on the other side of the rock outcrop? And even if they strongly preferred the western side, why did they not make more homogeneous use of that side's surfaces, rather than leaving some otherwise perfectly fine surfaces completely untouched while overcrowding others with petroglyphs?

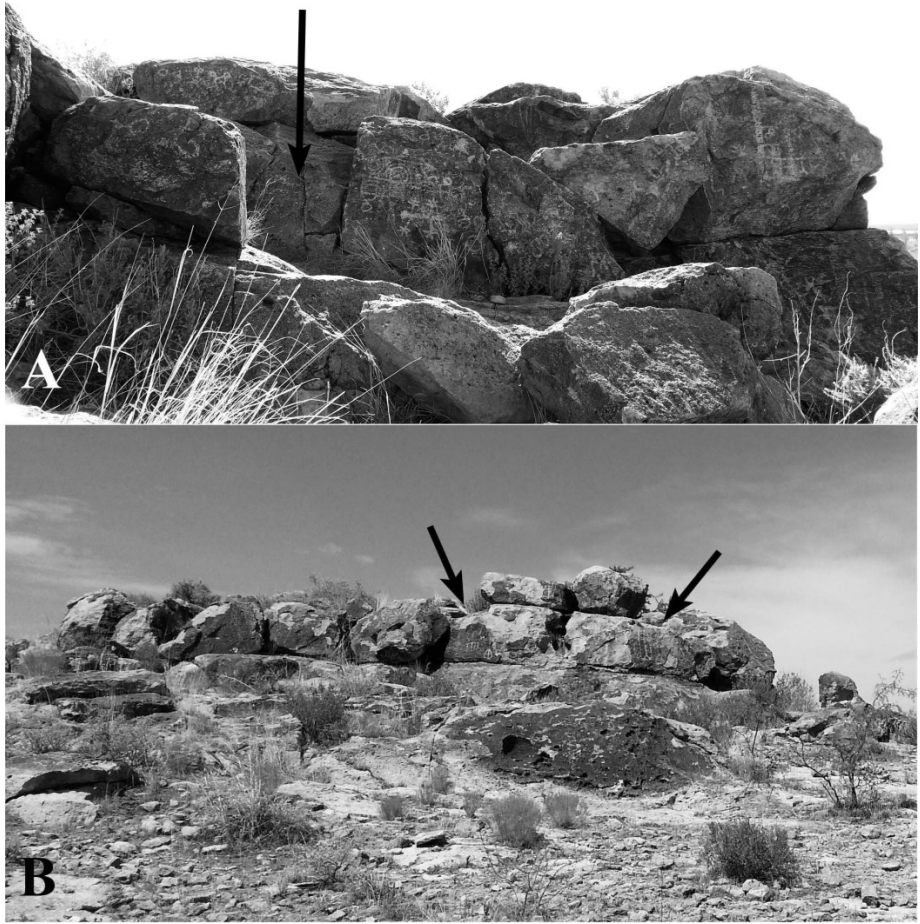

**Figure 1: Comparison of petroglyph manifestation on the west and east sides. A)** the west side of the *El Peñón del Diablo* has several densely clustered panels on it; arrow shows an inexplicable empty space between two of them, and **B)** the east side with only two panels, #70 and #72, indicated by arrows. (Photographs by Emiliano Gallaga)

This is a brief description of the rock art site called *El Peñón del Diablo*, located in the northern Mexican state of Chihuahua, which had its heyday in archaic times. But the site's uneven distribution of rock art motifs is just one illustrative and accessible example of a more general phenomenon: in many sites around the world the choices of where to make rock art seemed to have had little do with a concern for leaving visual representations for the benefit of other observers. For example, the same phenomenon has been observed in the context of rock art from Ice Age Europe. In the words of prominent French archaeologist Clottes:

> In Chauvet cave, the first chamber (Chamber of the Bear Wallows) is the largest in the cave and its walls, smooth and white, would seem a priori to be highly suitable. But paintings are present only right at the back. This came as such a surprise to me that, at the beginning of our research in the cave, I asked the geologists on the team whether some form of superficial degradation caused by natural phenomena (air currents, water flow, superficial calcification) might have occurred that could have destroyed any artwork. Their response ruled this out. (Clottes, 2016, p. 116)

Clottes suggests that the absence of images in the first chamber could be explained in terms of two principles. First, the artists might have preferred locations far away from the light of day, because the first paintings in Chauvet cave are found exactly at the edge of daylight, where one is already entering the permanent darkness of the cave but can still weakly perceive the light from the entranceway. Interestingly, this transition between light and darkness could have also played a role at *El Peñon*, given that the petroglyphs are concentrated on the western side of the outcrop, and thereby remain in the shadow of the first rays of the rising sun and, at the end of the day, receive the last rays of the setting sun.

Second, Clottes contends that the artists were not so much concerned with the technical suitability of a surface for producing a public image, but whether it personally attracted them or not: 'The artist needed to know whether this particular wall was suitable – not physically, but spiritually – for affixing an image and for what kind of image, or whether, on the contrary, the wall refused it or was devoid of power' (Clottes, 2016, p. 117). This materially mediated experience of what we may call a power spot would have been shaped by natural phenomena, e.g. Clottes highlights the role of enhanced acoustics and the presence of natural reliefs. But there will also surely have been a social dimension to the selection process, and particularly the presence of rock art left behind from previous cave visitors. Moreover, as Clottes' account indicates, such a power spot may have been perceived as having some kind of agency of its own, either attracting or refusing material engagement. This perspective may seem strange to us, but other cultures saw rock as potentially animate, sentient, and sacred (Dean, 2010). So, could it be that, rather than viewing the primary aim of such prehistoric rock art as the creation of images meant for public consumption, it is more appropriate to treat them as traces of materially mediated ritual interaction?

These two competing approaches have been actively debated over the last couple of decades in cognitive archaeology (e.g. Hodgson & Pettitt, 2018; Lewis-Williams, 2002; Malafouris, 2007). The field of cognitive archaeology tries to reconstruct the nature of the human mind in the deep past (Abramiuk, 2012), and two broad orientations can be distinguished in that field, which to a large extent mirror the divisions of philosophy of mind and cognitive science more broadly (Clark, 2014). We will refer to the traditional, dominant orientation as the *representational* approach, which holds that cognition essentially consists in the generation and manipulation of mental representations, such as updating internal world models, a process that is typically cashed out in computational, information-processing terms (Frith, 2007). Traditionally, the underlying operational basis of the mind is spatiotemporally restricted to the brain, and in this sense the representational approach is at the same time also an *internalist* approach; it limits the boundaries of the mind to the brain (Hohwy, 2013). In the context of cognitive archaeology, this approach is consequently interested in explaining the origins of the specifically human mind in terms of cognitive changes primarily driven by genetic evolution of the brain (e.g. Henshilwood & Dubreuil, 2011), with the implication that the archaeological artefactual record is relegated to a secondary role of being only an external cultural expression of these internal biological changes.

Here we will adopt an alternative, still relatively minor theoretical orientation, which is called the *enactive* approach. This approach started as a way of integrating phenomenology of embodiment into cognitive science (Thompson, 2007; Varela, Thompson, & Rosch, 2017), but its scope includes research of our situatedness in a sociocultural environment, including work in archaeology (e.g. Durt, Fuchs, & Tewes, 2017; Hutto, 2008; Malafouris, 2013). The foundational assumption of this approach is that embodied action in the world is not only a product, but also a constitutive part of cognitive processes, and this is also the case for culturally patterned forms of behavior (Di Paolo, Cuffari, & De Jaegher, 2018; Hutchins, 1995; Kirchhoff & Kiverstein, 2019). In other words, the basis of the mind is not restricted to an individual's brain; rather, it consists in the complex, self-organizing interactions of the individual's brain, body, and world, including of course interactions with artefacts and other agents (Di Paolo, Buhrmann, & Barandiaran, 2017). In this sense the enactive approach is at the same time also an *externalist*, or perhaps more adequately, a *relational* approach that adopts a systems perspective.

According to this approach, the origins of representations should not be explained by appealing only to a genetically primed brain, as if representations were biologically pre-built inside the brain, only waiting to be externalized after the emergence of an appropriate sociocultural context. Instead representations are enacted during the course of people's embodied interactions within a suitable sociocultural environment (Hutchins, 2010). In the context of cognitive archaeology, this approach is therefore interested in explaining how representational practices could have emerged from people's material engagement, and how the resulting representational practices could have in turn shaped the evolution of the human mind (Froese, 2019; Iliopoulos & Garofoli, 2016; Malafouris, 2007), eventually giving rise to the full range of symbolic cognition, including writing (Overmann, 2020; Stewart, 2010). Given that the starting point of the enactive approach is that the mind is, at its core, a self-organizing process of embodied interaction, it has found a natural ally in the *material engagement theory* developed in cognitive archaeology (Malafouris, 2013; Renfrew, 2004), which holds that our interaction with things shapes our mind. This integration of theoretical perspectives provides the background for this chapter's analysis of rock art. When applied to the origins of rock art in the Paleolithic, for example, it becomes possible to argue that when people's material engagement with cave walls left traces, these could then serve as the basis for the emergence of new forms of culturally mediated interaction, eventually scaffolding the development of specifically representational practices (Froese, 2019).

Yet it may be necessary for the enactive approach to zoom out its perspective even further: as this approach continues to develop outward from its theoretical roots in the biology of the organism-environment system, and hence to cast a broader view over all the complex networks of interactions that humans engage in, it seems fruitful for it to also ally with recent archaeological work on *material entanglement* (Hodder, 2012, 2018). This line of work has already attracted the interest of researchers in embedded, extended, and distributed cognition (Sutton, 2020; Wheeler, 2020), and there are potentially fruitful points of contact with the enactive approach to be explored as well. For instance, there is the key insight that spontaneously unfolding interactions can enable, yet also constrain, further actions, potentially resulting in "entrapment" (Hodder, 2012) or "bad habits" (Ramírez-Vizcaya & Froese, 2019). And there is also the hypothesis that over the long term complex networks of these interaction processes can add up to irreversible, self-amplifying processes that shape human evolution (Froese, 2018; Hodder, 2020). We hope that this chapter's contribution to research in material religion can serve as a step toward a deeper dialogue between these approaches.

**Toward an enactive approach to rock art**

The enactive approach has developed a distinct interpretive framework for the archaeological record of prehistoric artifacts (Garofoli, 2017; Malafouris, 2008), including rock art (Froese, 2019; Malafouris, 2007). Before we develop our case study of rock art, it is useful to get a better understanding of this approach to rock art by considering how it differs from the default assumptions of the representational approach.

    The representational approach, as the name suggests, tends to interpret the significance of rock art in representational terms. Thus, the intention of figurative imagery is often taken to be the creation of pictorial signs, be it of things perceived or imagined, while nonfigurative imagery tends to be treated as abstract signs. It is therefore not surprising that from this perspective the study of rock art is treated akin to the decipherment or reading of a prehistoric 'text': "researchers are readers of the art that has been left behind, those visual cues in many ways presenting themselves like graphic 'texts' made of strange, unknown signs" (David, 2017, p. 11). Moreover, analysis of the intentions and significance of the rock art is thereby *product*-oriented, i.e. biased to the static signs, while the active *process* of creating them is relegated to being just a contingent means of achieving that end. Consequently, in this traditional view rock art is also sharply distinguished from other forms of contingent substrate manipulation, such as the undulating lines found in some caves that were created by dragging fingers along a clay wall (cf. Clottes, 2016). And finally, the end goal of rock art is conceived of in the context of potential observers, who could in principle 'read' these signs. This representational approach to rock art as 'graphic texts' comes quite naturally to us modern observers, especially to literate academics, but it may over-intellectualize some forms of rock art.

    The enactive approach does not rule out that some rock art was created for specifically representational purposes, but it questions the validity of generalizing that purpose. Moreover, the metaphor of a graphic text does not fit very well with the archaeological record, especially regarding rock art from Paleolithic caves and certain hunter-gatherer contexts. We know that in many cases observability by others was not a primary concern of the artists, whereas a lot of attention was paid to the affordances for material engagement (Robert, 2017). This is reflected by a preference for suggestive rock forms, suitable substrates, and strenuous and messy access routes (Clottes, 2016; Hodgson & Pettitt, 2018; Lewis-Williams, 2002). In this view, there is a complex developmental continuity from the earliest forms of material engagement, such as getting one's hands dirty by leaving finger traces on cave walls, to the appearance of hand stencils and figurative cave art (Froese, 2019). In all of these cases a concern for embodied

interaction with the cave environment dominated, with the main difference being that eventually this concrete interaction process started to incorporate more conventional forms into the material engagement. Thus, especially when trying to understand the earliest traces of cultural expression of a region, rock art may be more appropriately conceptualized as concrete traces of an interaction process involving graphical acts, rather than as abstract graphical signs that were made in order to be read like a text (Malafouris, 2007).

     As such, the perceived meaning of such traces to prehistoric observers may have had more to do with how they were transforming the surface, for example by marking it as having participated in the realization of certain kinds of materially mediated rituals, sometimes even echoing the precise movements that were used. For example, a zigzag line on a cave wall is a material trace left behind by a person's zigzag movement, and the surface can be thought of as a material interface affording such kinds of patterned interactions. While it is possible to interpret this trace in terms of the artist's intention to leave a sign with a particular meaning, say a 'water snake' signaling the presence of water to others, we should not ignore the importance of the act itself. Arguably, what also could have mattered to the maker of the sign was the concrete experience of being in that time and place, and personally engaging with the surface, especially if it was a culturally recognized special place. We could even go as far as to say that it mattered less that the graphical patterns also had a representational meaning compared to the brute existential fact of the performance of its making, whereby the mark maker could come into direct contact with the powerful forces that animated the place. The bodily performance of a culturally significant pattern during this concrete material engagement, expressed by the arm and hand movements that embody the pattern in contact with the surface and record it for posterity as a graphical trace, can then be understood as a symbolically mediated way of intensifying this encounter between self and other. An illustration of this hypothetical example is provided in **Figure 2**.

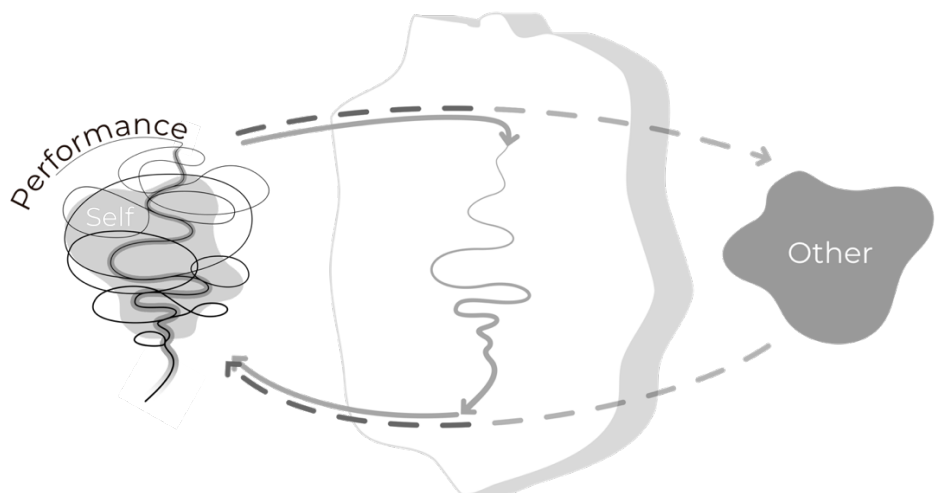

**Figure 2. An enactive approach to prehistoric rock art.** The artist intends to interact ritually with the powers that are animating a special place, including by manipulating its surface, and this materially mediated aspect of the performance remains visible as traces on the surface. Observers of these traces can see in them an encounter between Self and "Other," especially if the traces include culturally meaningful patterns that highlight this intention of the engagement. In this way the traces transform the appearance of the surface, making it stand out from other similar surfaces, and increasing its attraction for future ritual interactions. (Diagram by Maria Gohlke)

*A methodological challenge*

It is methodologically difficult to scientifically arbitrate between the representational and enactive approaches to prehistoric rock art. So far there has been a concern with qualitative hypotheses (Froese, 2019; Malafouris, 2007), as the enactive approach gives rise to a number of general expectations regarding the earliest forms of rock art:

- Simple marks and geometric motifs should predominate, and these should have been made without concern for their precise execution (e.g. given that the emphasis is on performing the interaction with the surface, rather than on its resulting material trace, we should not expect drawings to be improved).
- If figurative motifs are present, they should not show much concern for realism.
- The location of the marks should show a concern for material engagement with the rock's surface (e.g. by affording a particular kind of interaction, such as completing naturally suggestive contours or guiding placement of hand stencils).
- The location of the marks should show a concern for a special place that facilitates interaction with what lives in or lies behind the rock (e.g. by literally revealing the rock's interior or by being particularly expressive of its inner character).
- The location of the marks should have been chosen with little concern for access or enhanced visibility to external observers.
- When there is concern for the trace itself, it is mainly with the aim of enhancing the potency or duration of the interaction, as exemplified by hand stencils which first enhance the resemblance of hand and rock by covering them both with paint and subsequently leave a permanent echo of that direct contact.

Most of these criteria for the earliest forms of rock art have already been discussed in the literature, especially in the context of the shamanic hypothesis (Clottes, 2016; Clottes & Lewis-Williams, 1998; Lewis-Williams, 2002) and with respect to altered states of consciousness more broadly (Froese, 2013, 2015; Froese, Woodward, & Ikegami, 2013). Even those rock art researchers who prefer not to speculate about the prevalence of shamanistic cultures in deep prehistory, and who hence do not appeal to the extreme altered states of consciousness assumed by the shamanic hypothesis, are in broad agreement with these qualitative criteria (Hodgson & Pettitt, 2018). Nevertheless, the importance of material engagement, and of the entanglement to which it gives rise, has been difficult to measure and quantify.

     Here we propose a novel testing method: we argue that the distributions of rock art at a site should reflect whether the artists' intention was focused on the representational *product* of their activity or on the *process* of material engagement itself. The representational approach would expect rock art to be equally spatially distributed across surfaces, like the paintings on the walls of a gallery or text on a page, because the artists were concerned with making the traces more visible or 'readable'. They would avoid compromising public visibility of an image due to clutter and overlap with other images. However, to the extent that the rock art consists of traces of ritual performances, and that these can serve as signs of previous successful

interactions with the sacred via a specific surface of a special place, we can expect that such traces actually get entangled with subsequent performances at the same location. Hence, the enactive approach, and its claim that material engagement with such surfaces was more important than the artistic outcome per se, leads us to expect a skewed distribution whereby some areas are more crowded with rock art. There is an accrual of sacred power at particular surfaces because of this preferential attachment. In other words, we expect a self-amplifying effect, akin to the "Matthew effect" (Perc, 2014), which is the phenomenon that the rich tend to get richer and the potent even more powerful, an effect that is also studied in sociology in terms of cumulative advantage and success-breeds-success. This phenomenon has been intensely studied by network science since the 90s (Barabási, 2002; Watts, 2003), and we now know that this kind of skewed distribution, known as a "power law" to refer to its exponential change, tends to consistently emerge in social systems where people select among a broad range of choices according to their preferences. To test this methodological proposal, we analyzed the petroglyph distribution at *El Peñón del Diablo*, Chihuahua, Mexico.

**Description of *El Peñón del Diablo* site and petroglyphs**

In mid-2014, a series of meetings was held between the authorities of the Municipality of Janos and the staff of the *Instituto Nacional de Antropología e Hístoria* (INAH) Chihuahua Center with the aim of supporting the management of the municipality's cultural heritage.  Among the points that were addressed, was a request of support for recording and research at the site known as *El Peñón del Diablo* (The Devil's Crag). Due to the site's proximity to the community, it is a recreation area that is visited and used regularly, so there is neither adequate protection nor control of access, resulting in it being exposed to vandalism. Similarly, by being located within areas of cultivation and livestock breeding, the site is at risk of being lost little by little. In response to these risks, the Municipality of Janos requested the support of archeologists from the School of Anthropology and History of Northern Mexico (EAHNM), Chihuahua, to record the petroglyphs and survey the surrounding site in order to better understand and communicate its importance and contribute to its conservation and promotion as a patrimonial asset of the community (Gallaga & García, 2019).

The site *El Peñón del Diablo* is located in the Northwestern portion of the state of Chihuahua, within the municipality of Janos, just 2.6 kilometers southwest of the community of the same name, at an altitude of 1,366 meters above sea level (**Figure 3**). The locale, from which the site receives its name, is a natural outcrop of crystalline tuff that stands out in the valley on which a series of images were made, one of which is an anthropomorphic character with horns on the head that stands out and is locally known as '*el diablo'* (Spanish for 'the devil') (**Figure 4**). The *Peñón* is approximately 18 meters high with a northwest-southeast orientation and a length of 66 meters (**Figure 5**). It is located in the geographical region known as the intermountain valleys of Janos and is characterized by grasslands, xerophilous scrub, oak and coniferous forests (CONABIO, 2014; CONANP, 2006). The location of the site should not surprise us, since it is located at the center of a fertile valley in which there was a spring, which is nowadays dry possibly due to overexploitation of the aquifer. Only in the rainy season is the old riverbed filled with water and the valley comes back to life, attracting all the desert wildlife. The site is about 5 km east of the Casas Grandes River and the Cerro Juanaqueña site (a relevant archaic site) (Hard & Roney, 1998), and the later site of Paquimé is 60 km to the south (Gallaga & García, 2019; Gallaga, Moreno Alvarado, & Guzmán Aguirre, 2016).

The petroglyphs of *El Peñón del Diablo* were made by a combination of techniques between scraping and crushing/beating, since their designs are not very deep in the rock (i.e. they did not leave a well-marked groove) but only broke the surface layer of the rock. At the end of our field season a total of 78 panels were identified at *El Peñón del Diablo,* each consisting of one or more rocks, and a total of 651 petroglyphs were identified. Each one of them was recorded, drawn, and photographed. As we mentioned, this data was recorded as part of an archaeological survey of the area, and therefore it was not influenced by the current proposal. Although there is a presence of petroglyphs around all sides of the *Peñón*, most of them are concentrated on the western side, perhaps because that side is it is shadowed in the morning or well-lit at dusk at a time when ritual activity tends to begin. In addition, some of the placements are related to sunlight shining through cracks in the *Peñón* during sunrise on special days marked by astronomical events, including solstices (Muñoz, 2017).

Subsequently, the petroglyphs were categorized through the framework established by Viramontes (2005) resulting in 231 figurative motifs and 420 geometric motifs (**Figure 6**). Anthropomorphic (145) and zoomorphic (68) figures predominate the first category, while in the second category, lines (160), circles (99) and triangles (29) were the most common designs. On top of this, three different rock art styles have been identified: the Desert Abstract (Archaic period 8,500/5,500 BCE to CE 100/200), the Jornada Mogollon area (Archaic period 8,500/5,500 BCE to CE 100/200), and Medio Period (Paquime ceramic period CE 1,150-1,450). The Desert Abstract style is associated with the Archaic period and is distinguished by motifs such as zigzag lines, wavy lines, concentric circles, rake shapes, simple circles with dots and suns with rays, as well as hand and footprints, traces of animals and anthropomorphic figures (Schaafsma, 1980, 1992; VanPool, Rakita, & VanPool, 2009, p. 53). The Jornada Mogollon style is also associated with the Archaic period and is characterized by its naturalistic character and anthropomorphic representations of "hunting scenes, with animals such as mountain sheep or deer, and humans with headdresses who have shaman like qualities" (Sutherland, 2006, p. 9). The motifs identified for the Medio Period (CE 1,150-1,450) are lizards, snakes, tadpoles, outlines of crosses, circles with rays, anthropomorphic figures with horns and carrying a staff, as well as human figures with expressive arms and legs (Schaafsma, 1980, 1992, 2005; VanPool et al., 2009, p. 54). The Desert Abstract style was by far the most prevalent at the site, and we therefore do not further distinguish between these petroglyph styles in what follows.

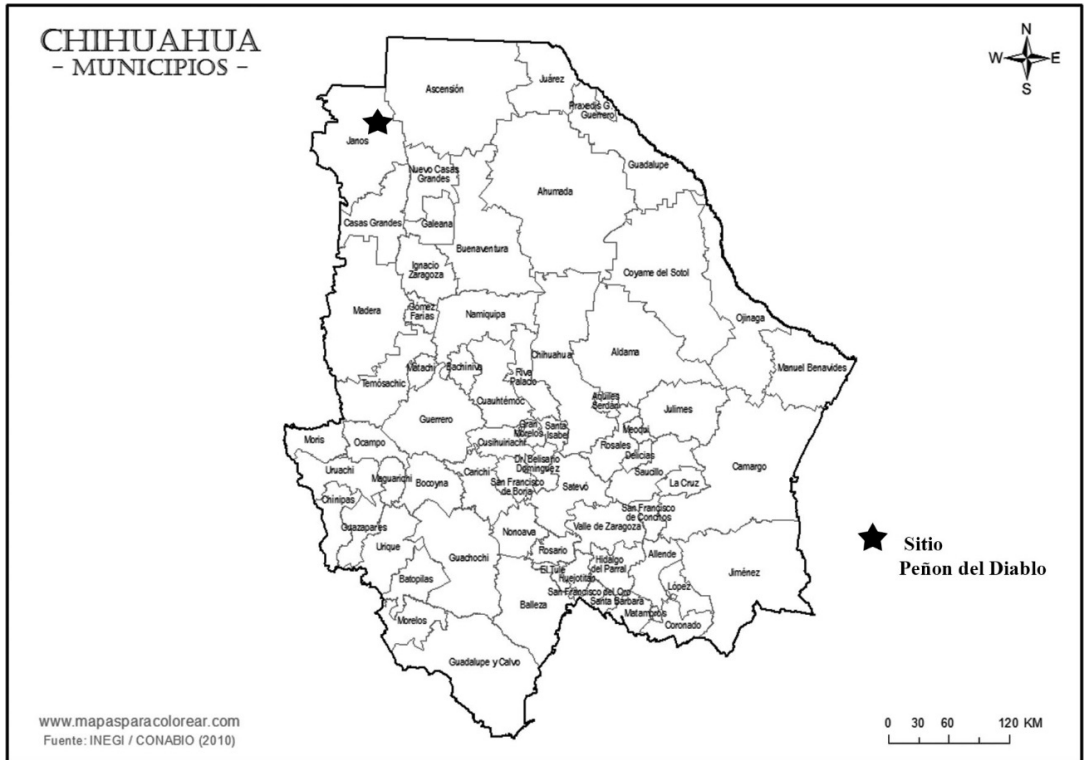

**Figure 3: Location of *Peñón del Diablo* site in the state of Chihuahua, Mexico.** The site is located in a municipality bordering the state of New Mexico, USA. (Map modified by Emiliano Gallaga)

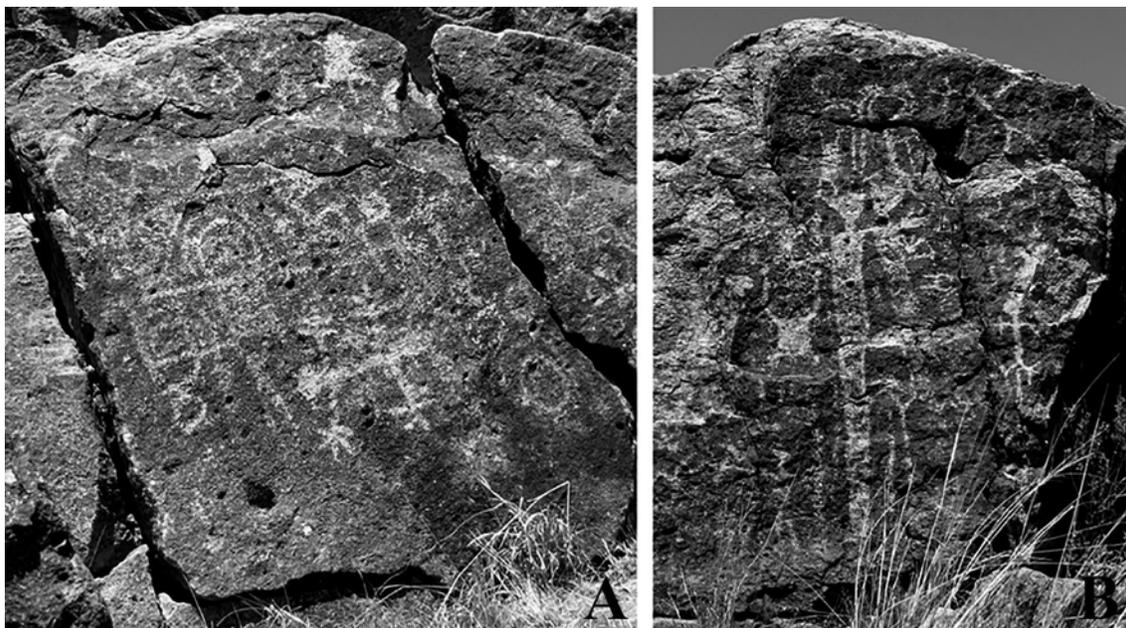

**Figure 4: Some illustrative examples of the petroglyphs at the *El Peñón del Diablo* site. A)** Panel #25, a surface densely clustered with petroglyphs; **B)** Panel #18, where the figure locally known as '*El Diablo*' is found. (Photographs by Emiliano Gallaga)

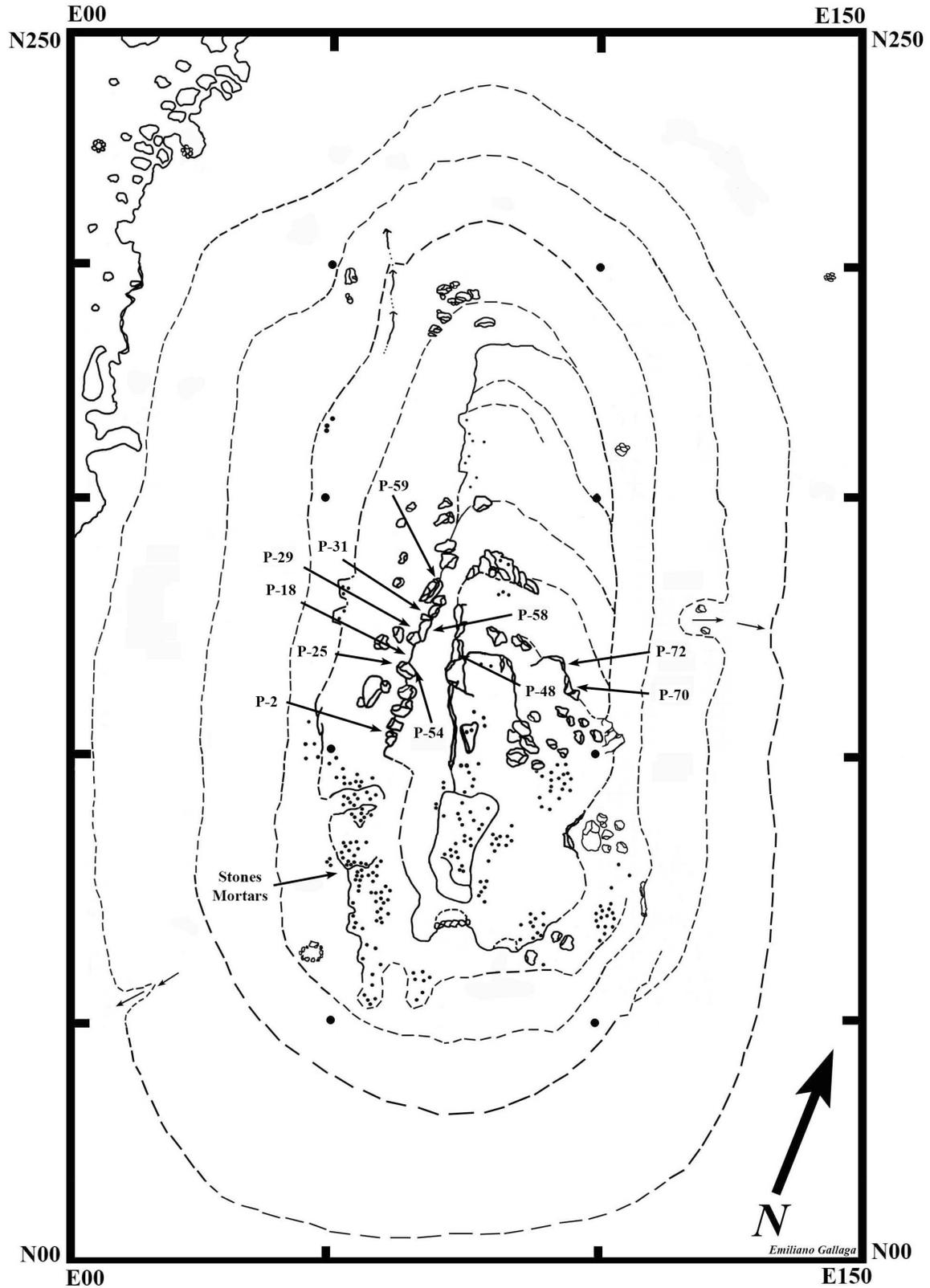

**Figure 5: *El Peñón del Diablo* map.** Details discussed in the text are marked in the map. (Map drawn by Emiliano Gallaga)

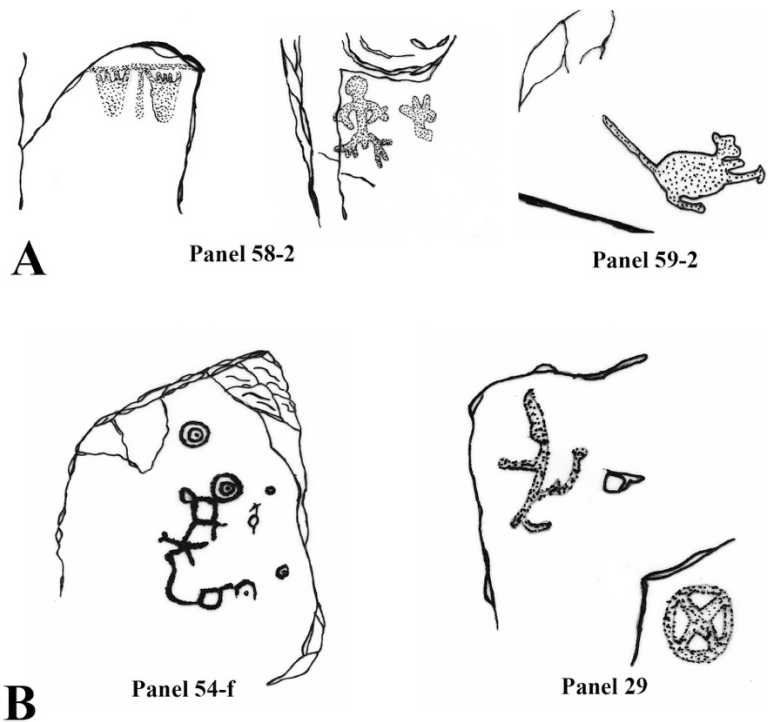

**Figure 6: Some examples of *El Peñón* petroglyphs: A)** Figurative petroglyph examples: panel 58-2 shows motifs from the desert abstract style, while panel 59-2 shows a rare example of what seems to be a cat. **B)** Non-figurative and geometrical petroglyph examples of the desert abstract style from two panels. (Drawings by Emiliano Gallaga)

*Archaeological context: Life around El Peñón del Diablo*

Investigations at rock art sites often only focus on the rock art and the rest of the site/area is left unexplored. Yet the rock art may well be part of a broader set of elements and materials that could indicate other types of activities which were also carried out by the community that made the petroglyphs or that occupied/used the areas surrounding them. In the present case, it was already known thanks to previous investigations that the site was much more than simply a rock art site. At its base, more than 200 fixed mortars were documented, which have been interpreted as 'the center of an extensive mesquite pod processing site during the Archaic Period' (VanPool, Rakita, & VanPool, 2015, p. 145).

Accordingly, the survey also included surface collections in two phases: one total collection in a grid of 250 meters x 150 meters, and another, less systematic prospection was made over the surrounding area of the *El Peñón* site using transects set 20-30 meters apart. The primary grid yielded 13,137 lithic artifacts (216.127 kg). The lithic sample consisted of 12,656 flakes, of which 267 (215 g) were obsidian, and 531 cores, 253 artifacts, and 228 projectile points. Preliminary results indicate that materials were being prepared at the *Peñón*, but the vast majority of the materials was arriving pre-worked from another location, probably at the raw material sources. So far, seven types of projectile point have been identified, most of which have a chronological distribution that places them between the Middle (3,000-1,000 BCE) and Late Archaic Periods (1,000 BCE to CE 200). In addition, a total of 25 ovens (identified as concentrations of fire-cracked rock on the surface), five possible structures, and an eroded human burial were registered, but no household in the site (Gallaga and García 2019).

Due to its location and time period, the *El Peñon del Diablo* site would have a strong interaction with the *Cerro Juanaqueña* and related 'Cerro de trincheras' (terraced hill) sites that flourished in the hills around the valley during the Late Archaic period around 3,000 years before the present (Hard & Roney, 1998). The massive *Cerro Juanaqueña* site is envisioned as a primary base camp mostly used for sleeping, and some item production, while other activities were performed in the valley. It is possible that the *El Peñón del Diablo* site was one of the many areas the Juanaqueña people used for ceremonies, social gathering, food consumption, and other social activities. As such, it would show us an important part of the multiple activities carried out by the inhabitants of this community outside the hill, as other researchers have already mentioned (Gallaga & García, 2019; VanPool et al., 2009).

The material evidence found around the site supports the claim that *El Peñón del Diablo* was not an isolated site in this part of the valley, but rather part of a larger complex of human presence and activity. In addition to the hundreds of petroglyphs, a large amount of mortars was found around the outcrop. These mortars were slowly carved into the bedrock through repeated grinding and indicate the presence of large numbers of people repeatedly returning to this site over long stretches of time. Thus, at first glance, one could think that it is only a place of ceremonial or ritual use; however, the investigations carried out both at the *Peñón* and in the surrounding area allow us to come to a different conclusion. Surely there was a ritual use or meaning of the *Peñón*; however, the site was also an area that was occupied, at least temporarily or seasonally. It also shows a great diversity of activities that transformed the natural environment into a cultural landscape recognized by communities that lived in the area for many generations. Accordingly, the activities surrounding the petroglyphs could have been of a communal nature since there is no impediment for them to have been observed by the people temporarily settled at the site (Bech, 2017; Muñoz, 2017; Murray, 2007; Rodríguez, 2016). It is plausible that the *Peñón* site served as a special place, that is, a public place at which people from different terraced hill communities in the wider area would gather for ritual, exchange, and other social activities on solstices (Gallaga and García 2019). This phenomenon of special places serving to integrate broad social networks is consistently found throughout the world and is associated with the rise of stateless social complexity (Stanish, 2017).

In sum, we can assume that the activities resulting in this corpus of petroglyphs were diverse. In this region of the Chihuahuan desert, a spring/waterhole would have been a highly attractive, life-giving feature, which presumably also imbued the location with a magical connotation. Accordingly, many of the petroglyphs were most likely associated with ceremonial/ritual activities, some of which were probably astronomical in nature. Others were probably associated with cultural or community affiliation, and/or claiming the place as theirs (Gallaga & García, 2019). This last claim is supported by the fact that not all of the cultural groups that frequented the area left their distinctive rock art at the site, such as the Apaches. Unfortunately, for us, there is no longer a living community that can claim ties to the archaeological remains of *El Peñon del Diablo.* As an archaic hunter-gatherer group that was experimenting with agriculture, they possibly transformed themselves into a sedentary agricultural community, which with the passage of time they could have become part of the inhabitants of the nearby Paquime site of the Medio period (CE 1,150-1,450).

This archaeological survey allows us to paint a richer picture of life around the *Peñón*. When you are in the highest part of the rock outcrop, your view perfectly dominates the valley around it. With a little imagination, it is not very difficult to glimpse it as it could have looked on a day in the Archaic period. The site would have been blessed with water from the nearby marsh, attracting birds and turtles, and of course people, some of whom would be crushing the seedpods of the nearby mesquite trees in the mortars that are at the base of the *Peñón*, while others worked stones to make projectile points or other needed stone tools. Looking past the marsh a little further south, smoke would be laying over the area where another group of people were sitting around the ovens preparing food, such as agave hearts, juicy portions of deer meat or rabbit, all of which would be eaten following the next solstice ceremony. Perhaps it was during such ceremonies when the communities' spiritual leaders would intend to interact with the divine forces of nature that reside in the site, and thereby leave traces of new images on the walls of the crag of *El Diablo*.

**Methods**

We propose that one way to test the enactive approach is to measure the distribution of rock art and examine whether it is skewed in accordance with a power law. Effectively, we argue that if the material engagement with special places is important, then some surfaces will tend to attract more people to connect with them, which in turn makes it even more likely to attract more people, and so forth. If this appeal to preferential attachment is on the right track, then we can adopt measures from network science that are designed to capture the popularity of people or things in terms of the skewed distribution of the numbers of their social connections or followers. To give just one contemporary example, this effect has been observed in the evolution of the structure of the Internet (Huberman, 2001), but also in more specialized applications such as in recent social tagging services (Hashimoto, Oka, & Ikegami, 2017).

If preferential attachment is at play in a social network, the prediction is that there should be an inverse correlation between the number of connections and the number of nodes with that number of connections: there is a small probability for a node to have a large number of connections, while there is an increasingly higher probability for a node to have an increasingly smaller number of connections. For our purposes, we treated the panels identified by the archaeological survey as the nodes, while the petroglyphs are the connections. For the survey some panels had been divided into subpanels based on different features, but we treated them as a single rock panel for our analysis.

We performed our analysis according to the following procedure: for each possible number of petroglyphs *k* on a panel's surface, which for this site was within the range *k* = [1, 25], we counted the number of panels whose surface had a particular number of petroglyphs $n_k$. This allowed us to calculate the proportion of those particular panels with respect to the total amount of panels *N*. We denote this proportion as the probability of a panel surface having a particular number of petroglyphs, P(*k*), which is given by the following equation:

$$P(k) = n_k / N$$

A secondary aim of our analysis was to take into consideration the specific kind of petroglyph forms associated with the panels. This is not the place to do a systematic analysis of all the petroglyph forms, but we can usefully distinguish between figurative motifs and all other, non-figurative (including non-identifiable) motifs. This distinction allows us to test whether there is a bias toward non-figurative forms for the panels containing the largest number of petroglyphs (large $k$). For if it is indeed the case that more densely worked panels were particularly attractive because of their potential for material engagement, then we might expect more simple graphic elements than elaborated figurative images.

We therefore calculated the average percentage of figurative motifs for each panel petroglyph cluster size $k$, with the expectation that there will be a decrease of the average percentage of figurative motifs with increasing cluster size.

**Results**

If the artists had distributed the petroglyphs in a more or less equally spaced manner across these panels, then we would have found the distribution of petroglyphs centered on an average of around 10 petroglyphs per panel. This is not what we found. The distribution of petroglyphs is skewed in the form of a power law, as we had expected, with the highest proportion of surfaces featuring very small numbers of petroglyphs, followed by a quick drop in proportion, and a long tail toward increasingly lower proportions of surfaces with higher numbers of petroglyphs. Some illustrative examples of different cluster sizes are shown in **Figure 7**, while the main results of our analysis are summarized in **Figure 8A**.

To our knowledge this is the first time that it has been reported that a petroglyph distribution is consistent with an explanation centered around the social phenomenon of preferential attachment. Admittedly, the fit to the power law curve could be better, but the overall trend is still very suggestive, especially given that the *Peñón* has less than 100 panels, whereas studies of modern social phenomena typically work with cleaner datasets that are larger by several orders of magnitude.

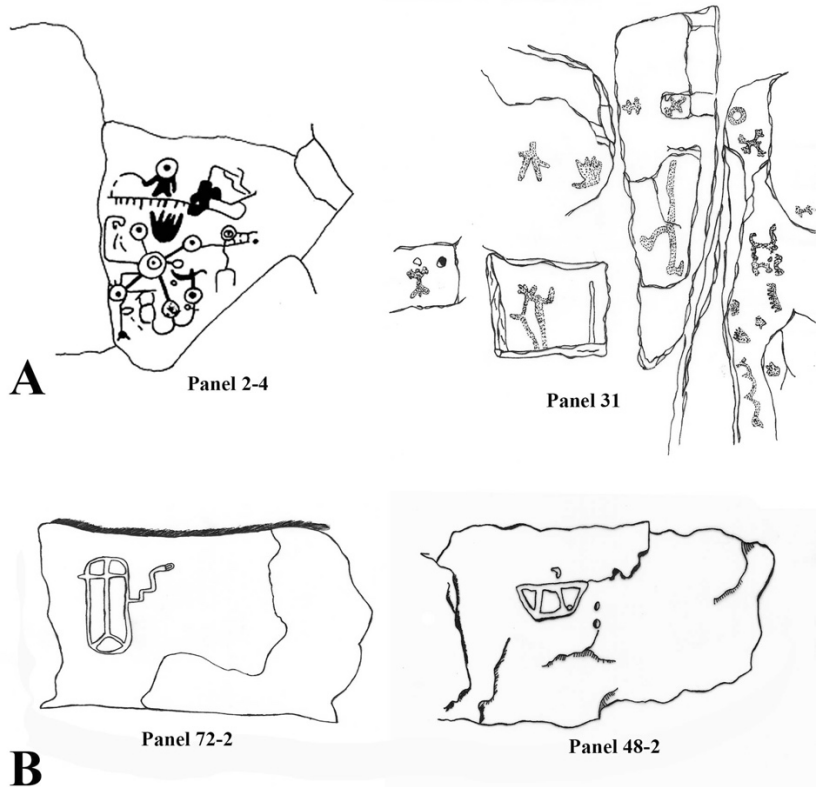

**Figure 7: Some examples of the distribution of the panels:** A) rock panels with several petroglyphs in a cluster, B) rock panels with isolated motives. (Drawings by Emiliano Gallaga)

Regarding the distribution of rock art figurative compared to non-figurative motifs, after averaging across all panels of a particular cluster size, we found that the average percentage of figurative motifs is only 38%. In other words, most motifs are non-figurative, which is consistent with our general expectation that there was more of a concern with material engagement, and hence comparably less concern with elaborating figurative motifs. However, surprisingly, we were unable to confirm our expectation of a decrease in the percentage of figurative motifs with respect to increased petroglyph cluster size: as shown **Figure 8B**, the average percentage of 38% is notably independent of cluster size. There is more variation for smaller cluster sizes, but this is expected given that fluctuations around the average will have a comparably larger effect when the total number of motifs is smaller.

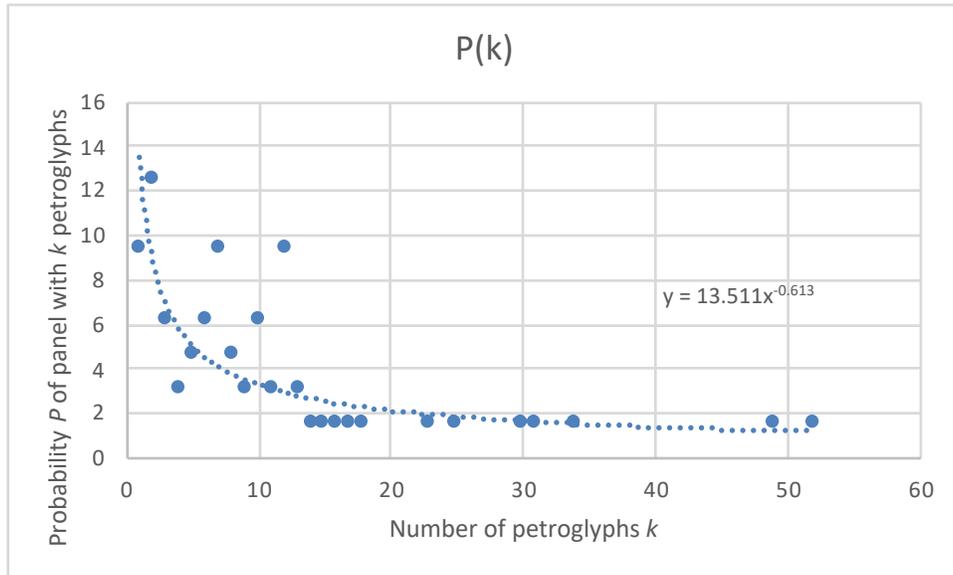

A

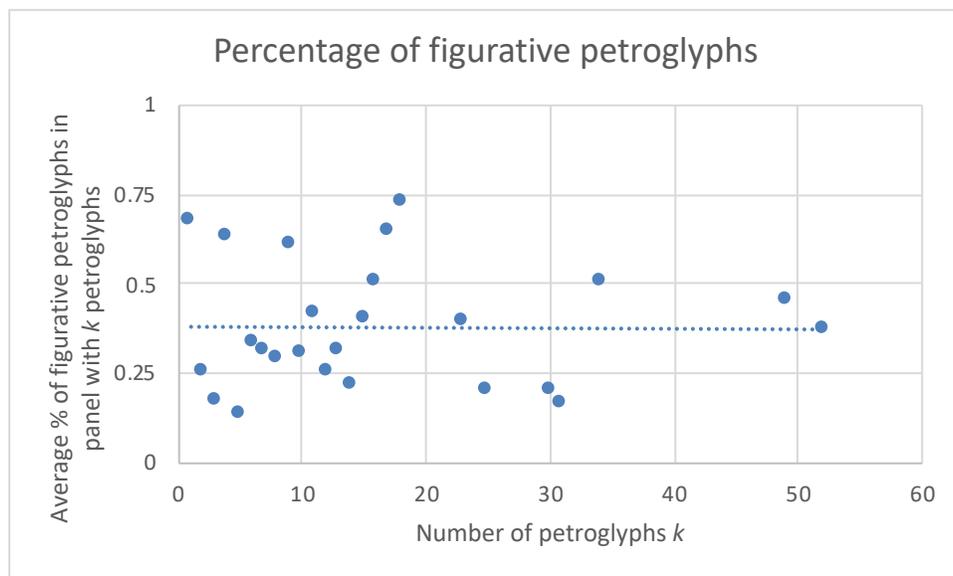

B

**Figure 8. Distributions of petroglyphs. A:** Scatter plot of the probability *P* of a rock panel with a particular number of petroglyphs *k*. Most rock panel surfaces have very few petroglyphs, while very few surfaces contain many petroglyphs. **B:** Scatter plot of the average percentage of petroglyphs consisting in figurative motifs for a rock panel with a particular number of petroglyphs *k*. Overall, 38% of petroglyphs are figurative, and this average is independent of petroglyph cluster size *k*.

What could explain this independence of the ratio of figurative to non-figurative motifs with respect to cluster size? It could indicate that these two kinds of motifs were actually not significantly distinct categories from the perspective of the artists. Thus, perhaps even the figurative motifs should be treated as equally part of the performance. For instance, in the case of figurative paintings in the Paleolithic caves of Europe Lewis-Williams (2002, p. 193) has questioned the importance of representational reference: "In all probability the makers did not suppose that they 'stood for' real animals." In addition, the solidity of the rock panels may have also played a role, given that material engagement only resulted in traces if the artists used tools with the aim of penetrating the surface of the rock. Given this material barrier to making rock art, any extra effort of elaborating complex figures rather than simple marks may be negligible. It would be interesting to do a comparative analysis to see if the ratio changes across cluster size if the substrate is softer, like clay walls.

**Discussion**

The distribution of the petroglyph concentrations at the *Peñón* makes a relatively good fit with a power law, but the fit could be improved. Part of the problem is that this is a small site and so the rock art corpus we analyzed is still a comparatively small dataset, and so deviations of data points from the exact power law distribution are to be expected. Future work could apply this method to rock art datasets that are orders of magnitude larger than ours, which should hopefully provide a better fit. Moreover, in this exploratory analysis we worked with the raw counts of rock panel surfaces and petroglyphs, while ignoring other possible factors that could influence the distribution. For instance, thematic relatedness could explain some cases of co-occurrences of petroglyphs on one rock, while smaller rock surfaces and/or larger petroglyphs could tend to decrease co-occurrence. We believe that such factors did not influence our findings in any significant way, but future work could for instance divide the rock surface into equally spaced areas in order to eliminate variation in panel size.

    We do not know how many people were involved in the creation of this rock art. It would be interesting if we could assign each petroglyph to a particular maker, which would allow us to develop a more detailed account of the entanglement between artists and rocks. However, for present purposes this information about authorship is not essential. Our proposal is that the placement of the petroglyphs was primarily driven by processes akin to preferential attachment in the context of material engagement, rather than for representational purposes, and this proposal is not dependent on whether the distribution of petroglyphs reflects the preferences of a large group of people that each created only one or few petroglyphs, or of a select number of people that created multiple petroglyphs. Certainly, our proposal would be further strengthened if it could be shown that rocks with high densities of petroglyphs are the product of temporally extended processes, involving multiple people and perhaps even multiple generations. That this expectation is on the right track is indicated by the temporal distribution of the rock art. Most of it was made during the Archaic period (8,500/5,500 BCE to CE 100/200), but it also attracted artists during the later Medio Period (Paquime ceramic period CE 1,150-1,450). This is an interesting topic for future research.

    Our proposal could also be further strengthened by applying the method to a suitable control case. In particular, when applied to a set of visual signs that we know were mainly made for representational purposes, we should not find evidence of a distribution with a power law. This is intuitively true of modern contexts such as art galleries, which aim to show each artwork in a stand-alone manner and hence have a narrow range of densities of artworks. Nevertheless, this contrasting expectation should also be confirmed with rock art sites, especially if there are more recent ones for which we have good ethnographic information confirming that the artists' primary intent was to create public representations. In this way network science could help to

further investigate the relationship between material engagement and material entanglement in prehistoric rock art.